\documentclass[prl,twocolumn,aps,showpacs]{revtex4}

\usepackage{here}
\usepackage{graphicx}
\usepackage{wrapfig}
\newcommand{\be}{\begin{equation}}
\newcommand{\ee}{\end{equation}}

\newcommand\pictc[5]{\begin{figure}
                       \vspace*{-0mm}\centerline{\vspace*{0mm}
                       \includegraphics[width=#1\columnwidth]{#3}}
                       \protect\caption{\protect\label{fig:#4} #5}
                    \end{figure}\vspace*{0mm}            }

\newcommand\pict[4][1]{\pictc{#1}{!H}{#2}{#3}{#4}}
\newcommand\rpict[1]{\ref{fig:#1}}

\newcommand\leqt[1]{\protect\label{eq:#1}}
\newcommand\reqtn[1]{\ref{eq:#1}}
\newcommand\reqt[1]{(\reqtn{#1})}

\newcounter{Fig}

\begin{document}
\begin{sloppy}

\title{Slow-light optical bullets in arrays of nonlinear Bragg-grating waveguides}

\author{Andrey A. Sukhorukov}
\author{Yuri S. Kivshar}

\affiliation{Nonlinear Physics Centre and Centre for
Ultrahigh-bandwidth Devices for Optical Systems (CUDOS), Research
School of Physical Sciences and Engineering, Australian National
University, ACT 0200 Canberra, Australia}
\homepage{http://www.rsphysse.anu.edu.au/nonlinear}

\begin{abstract}
We demonstrate how to control independently both spatial and
temporal dynamics of slow light. We reveal that specially designed
nonlinear waveguide arrays with phase-shifted Bragg gratings
demonstrate the frequency-independent spatial diffraction near the
edge of the photonic bandgap, where the group velocity of light can
be strongly reduced. We show in numerical simulations that such
structures allow a great flexibility in designing and controlling
dispersion characteristics, and open a way for efficient
spatiotemporal self-trapping and the formation of slow-light optical
bullets.
\end{abstract}


\maketitle

Nonlinear response is the fundamental property  of optical materials
that leads to interaction of propagating optical waves and allows
all-optical switching for various applications. Efficiency of
nonlinear interactions can be greatly enhanced in the regime of
light propagation with small group
velocities~\cite{Soljacic:2002-2052:JOSB}. The slow-light regime can
be realized when the frequency is tuned close to the edge of a
photonic bandgap where optical waves experience strong Bragg
scattering from a periodically-modulated dielectric structure. These
effects were studied extensively in the structures where the
propagation direction is fixed by the waveguide geometry, including
experimental observations of pulse propagation in
fibers~\cite{Eggleton:1999-587:JOSB} and AlGaAs ridge
waveguides~\cite{Millar:1999-685:OL} with Bragg gratings, or
coupled-defect waveguides in photonic crystals with two- or
three-dimensional modulation of the refractive index~(see, e.g.,
Refs.~\cite{Vlasov:2005-65:NAT, Gersen:2005-73903:PRL,
Jacobsen:2005-7861:OE}). In particular, it was demonstrated that the
group velocity and the corresponding pulse delay can be tuned
all-optically in nonlinear photonic structures, and at the same time
nonlinearity may compensate for pulse broadening due to
dispersion~\cite{Eggleton:1999-587:JOSB}.

All-optical control of the spatial beam dynamics becomes possible in
periodic two- (2D) or three-dimensional (3D) photonic structures,
where light can propagate in various directions. The possibility for
beam steering is inherently linked to the effect of beam spreading
due to diffraction. Similar to pulses, nonlinearity can suppress
beam spreading and support the formation of non-diffracting spatial
optical solitons~\cite{Kivshar:2003:OpticalSolitons}. Recent studies
have emphasized many unique properties of spatial solitons in
periodic structures such as waveguide arrays or photonic
lattices~\cite{Christodoulides:2003-817:NAT,
Fleischer:2005-1780:OE}, where modulation in the transverse spatial
dimension fundamentally affects the nonlinear wave dynamics. There
appear multiple angular band gaps supporting the formation of
spatial gap solitons, which possess tunable steering
properties~\cite{Rosberg:2005-2293:OL}. Recent
experiments~\cite{Eisenberg:2001-43902:PRL} demonstrated the effect
of simultaneous spatial and temporal self-trapping of optical pulses
in the form of optical light bullets~\cite{Silberberg:1990-1282:OL}.

\pict{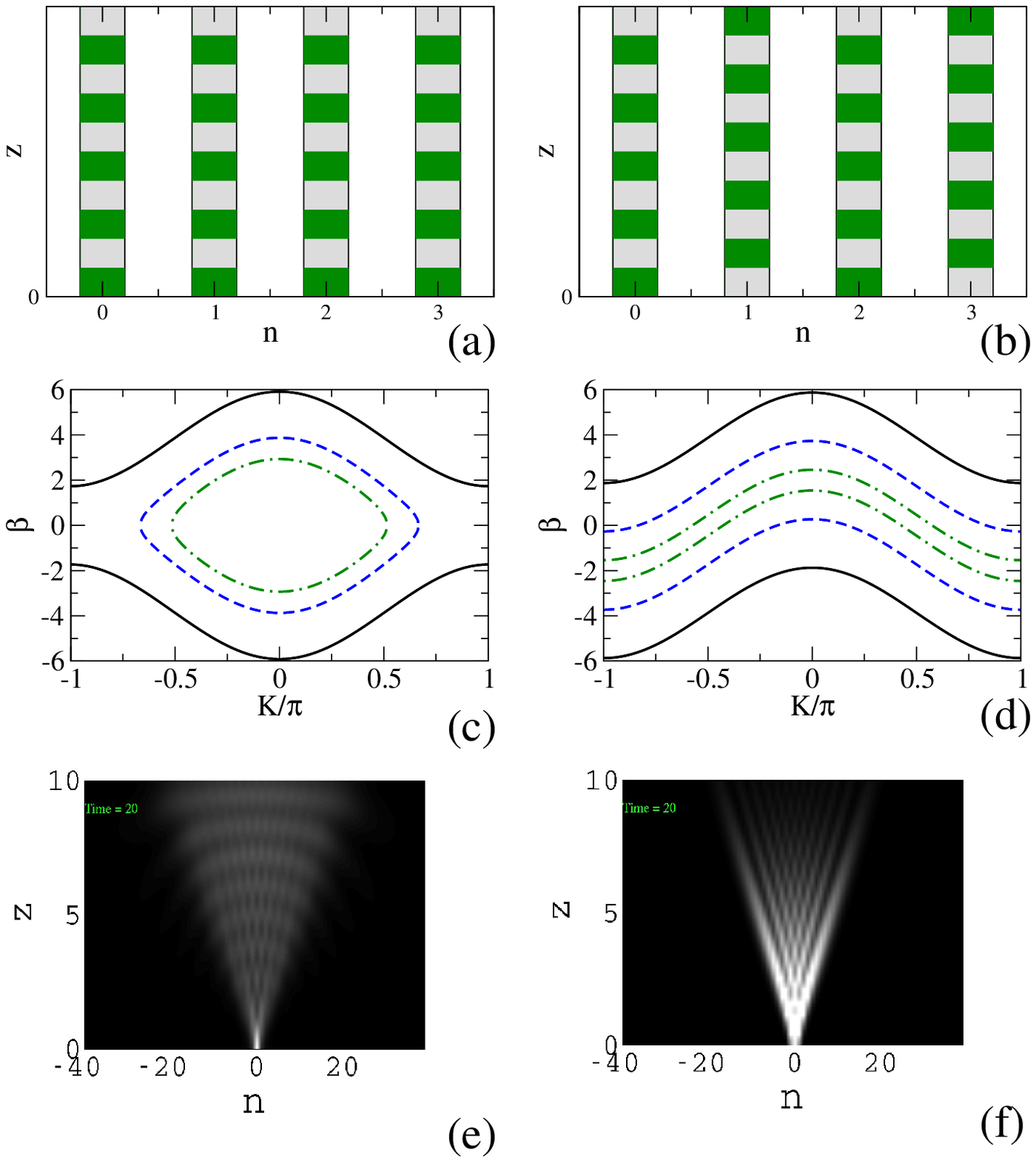}{array}{ (a,b)~Schematic of a waveguide array with
(a)~in-phase ($\varphi=0$) and (b)~phase-shifted ($\varphi=\pi$)
Bragg gratings; (c,d)~corresponding isofrequency contours for
different detuning from the gap edge: $\omega=4$ (solid), $\omega=2$
(dashed), $\omega=1.1$ (dashed-dotted). (e,f)~Corresponding
diffraction patterns for a pulse with the central frequency at
$\omega=1$. }

In this Letter, we address an outstanding key problem and
demonstrate how to perform dynamical tunability over both the
magnitude and direction of the speed of light through all-optical
control in the slow-light regime. In order to realize an effective
nonlinear control, it is necessary to balance the effects of
temporal dispersion in the propagation and spatial diffraction in
the transverse directions. For this purpose, instead of usually
studied symmetric photonic crystals~\cite{Altug:2005-111102:APL},
here we suggest to consider the photonic structures with
multiple-scale modulations in different directions. This overcomes
the limitations of the previously studied 2D or 3D defect-free
photonic-crystal structures where diffraction usually increases for
smaller group velocities due to a shrinkage of the isofrequency
contours~\cite{Joannopoulos:1995:PhotonicCrystals}, thus greatly
restricting the efficiency of spatial self-focusing of slow light.
Indeed, only quasi-localized nonlinear states with large spatial
extent were predicted for a slab 2D waveguide with a 1D Bragg
grating~\cite{Dohnal:2005-209:STAM}. In a sharp contrast, in this
Letter we show that it is possible to realize the {\em
frequency-independent diffraction near the edge of the bandgap} in
specially designed Bragg-grating waveguide arrays. This allows to
engineer independently the strength of diffraction and dispersion in
the slow-light regime, and thus provides optimal conditions for the
nonlinear control of spatiotemporal pulse dynamics.  In particular,
we predict and demonstrate numerically the formation of strongly
localized slow-light optical bullets in such structures.

The arrays of nonlinear optical waveguides which are homogeneous in
the propagation direction have been extensively studied in recent
years, and many possibilities for spatial beam control were
demonstrated experimentally~\cite{Christodoulides:2003-817:NAT,
Fleischer:2005-1780:OE}. We reveal that new regimes of the
spatiotemporal dynamics can be achieved in the waveguide arrays
modulated in the propagation direction with the period satisfying
the Bragg-resonance condition at the operating frequency. Such
structures can be fabricated in AlGaAs samples, with
accessible fast nonlinearity.

In the vicinity of the Bragg resonance, the evolution of optical
pulses can be modeled by a set of  coupled-mode nonlinear
equations~\cite{Agrawal:1989:NonlinearFiber} for the slowly varying
envelopes of the forward ($u_n$) and backward ($v_n$) propagating
fields in each of $n-$th waveguide. In the normalized form, these
equations can be written as follows,
\begin{equation} \leqt{CM}
   \begin{array}{l} {\displaystyle
      i \frac{\partial u_n}{\partial t}
      + i \frac{\partial u_n}{\partial z}
      + C ( u_{n-1} + u_{n+1} )
   } \\*[9pt] {\displaystyle \qquad \qquad
      + \rho_n v_n
      + \gamma (|u_n|^2 + 2 |v_n|^2) u_n
      = 0,
   } \\*[9pt] {\displaystyle
      i \frac{\partial v_n}{\partial t}
      - i \frac{\partial v_n}{\partial z}
      + C ( v_{n-1} + v_{n+1} )
   } \\*[9pt] {\displaystyle \qquad \qquad
      + \rho_n^\ast u_n
      + \gamma (|v_n|^2 + 2 |u_n|^2) v_n
      = 0,
   } \end{array}
\end{equation}
where $t$ and $z$ are the dimensionless time and propagation
distance, respectively, $C$ is the coupling coefficient for the
modes of the neighboring waveguides, $\rho_n$ characterizes the
efficiency of scattering from the Bragg grating, $\gamma$ is the
nonlinear coefficient, and the group velocity far from the Bragg
resonance is normalized to unity.

We reveal that both diffraction and dispersion can be precisely
tailored by introducing {\em a phase shift} between the otherwise
equivalent waveguide gratings, as illustrated in
Figs.~\rpict{array}(a,b). Only two- and three-waveguide nonlinear
couplers with in-phase gratings were analyzed
before~\cite{Mak:2004-66610:PRE, Gubeskys:2004-283:EPD}. Here, we
consider the effect of linear phase shift of the gratings across the
array characterized by the scattering coefficients $\rho_n = \rho
\exp(-i \varphi n)$. With no loss of generality, we can take $\rho$
to be real and positive. Then, the wave propagation in the linear
regime (at small intensities) can be fully defined through the
properties of Floquet-Bloch eigenmodes,
\begin{equation} \leqt{Bloch}
   \begin{array}{l} {\displaystyle
      u_n = u_0 \exp\left[ i K n + i \beta z - i \omega t \right] ,
   } \\*[9pt] {\displaystyle
      v_n = v_0 \exp\left[ i (K+\varphi) n + i \beta z
                                             - i \omega t \right] ,
   } \end{array}
\end{equation}
where $K$ and $\beta$ are the transverse and longitudinal components
of the Bloch wavevector. After substituting Eqs.~\reqt{Bloch} into
the linearized equations~\reqt{CM} (with $\gamma=0$), we obtain the
following relations,
\begin{equation} \leqt{mode}
   \begin{array}{l} {\displaystyle
      u_0 [ \omega-\beta + 2 C \cos(K)] + v_0 \rho = 0,
   } \\*[9pt] {\displaystyle
      u_0 \rho + v_0 [ \omega+\beta + 2 C \cos(K+\varphi)] = 0.
   } \end{array}
\end{equation}
These are the eigenmode equations which define the dispersion
properties of the Bloch waves, $\omega(K,\beta)$. Since
Eqs.~\reqt{mode} represent a square $2 \times 2$ matrix, there will
appear two branches of the dispersion curves. These dependencies
determine the key features of the wave spectrum. First, the gaps may
appear for a certain frequency range, where the propagating waves
with real $K$ and $\beta$ are absent. We notice that quasi-2D
spectral gaps can indeed appear for the modes localized in the
high-index waveguides, i.e. below the light-line of the substrate.
Second, the spatial beam refraction and diffraction are defined by
the shape of the corresponding isofrequency contours. The
propagation angle is defined by the value of $\alpha( \omega, K ) =
- \partial \beta /
\partial K$, and the effective diffraction experienced by the beam
depends on the curvature, $D( \omega, K ) = - \partial^2 \beta /
\partial K^2$. Below, we analyze how these fundamental characteristics
can be controlled by selecting the phase shift $\varphi$ between the
Bragg gratings.

For the in-phase gratings [i.e., when $\varphi=0$; see
Fig.~\rpict{array}(a)], the dispersion relation becomes
$\omega(K,\beta) = -2 C \cos(K) \pm (\rho^2+\beta^2)^{1/2}$. It
follows that the 2D gap appears only when the waveguide coupling is
weak, $C < \rho/2$. On the other hand, the shape of the isofrequency
contours is strongly frequency-dependent as the transmission band
edge is approached, see Fig.~\rpict{array}(c). This happens because
the position of the one-dimensional frequency gap depends on the
propagation direction.

\pict{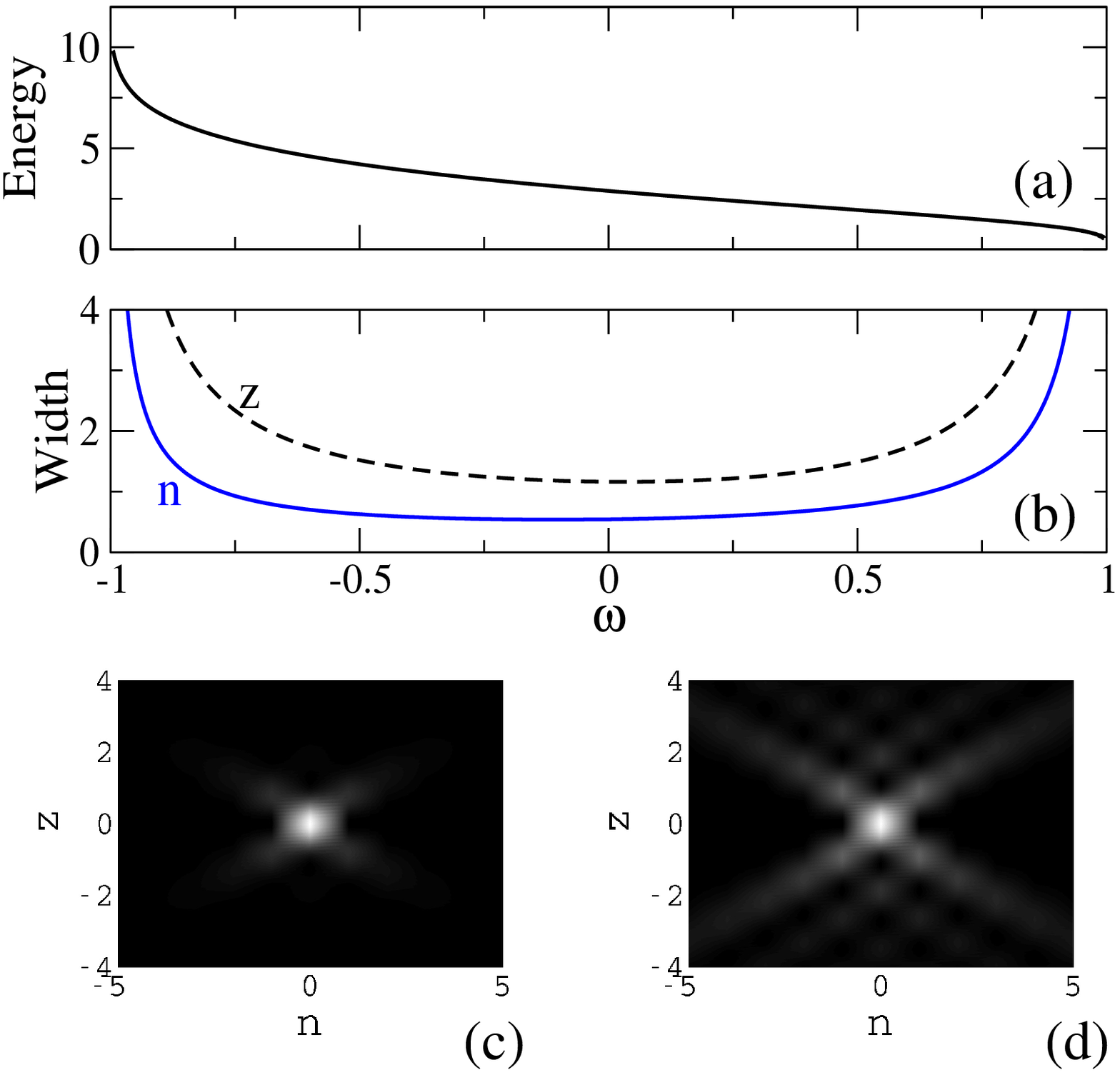}{familyFreq}{(colour online) Family of the
stationary (zero-velocity) light bullets characterized by (a)~energy
and (b)~width along the transverse (solid) and longitudinal (dashed)
directions vs. the frequency tuning inside the spectral gap.
(c-d)~Intensity profiles of slow-light optical bullets for
(c)~$\omega=0.8$ and (d)~$\omega=0.995$; brighter shading marks
higher intensity. }

\pict{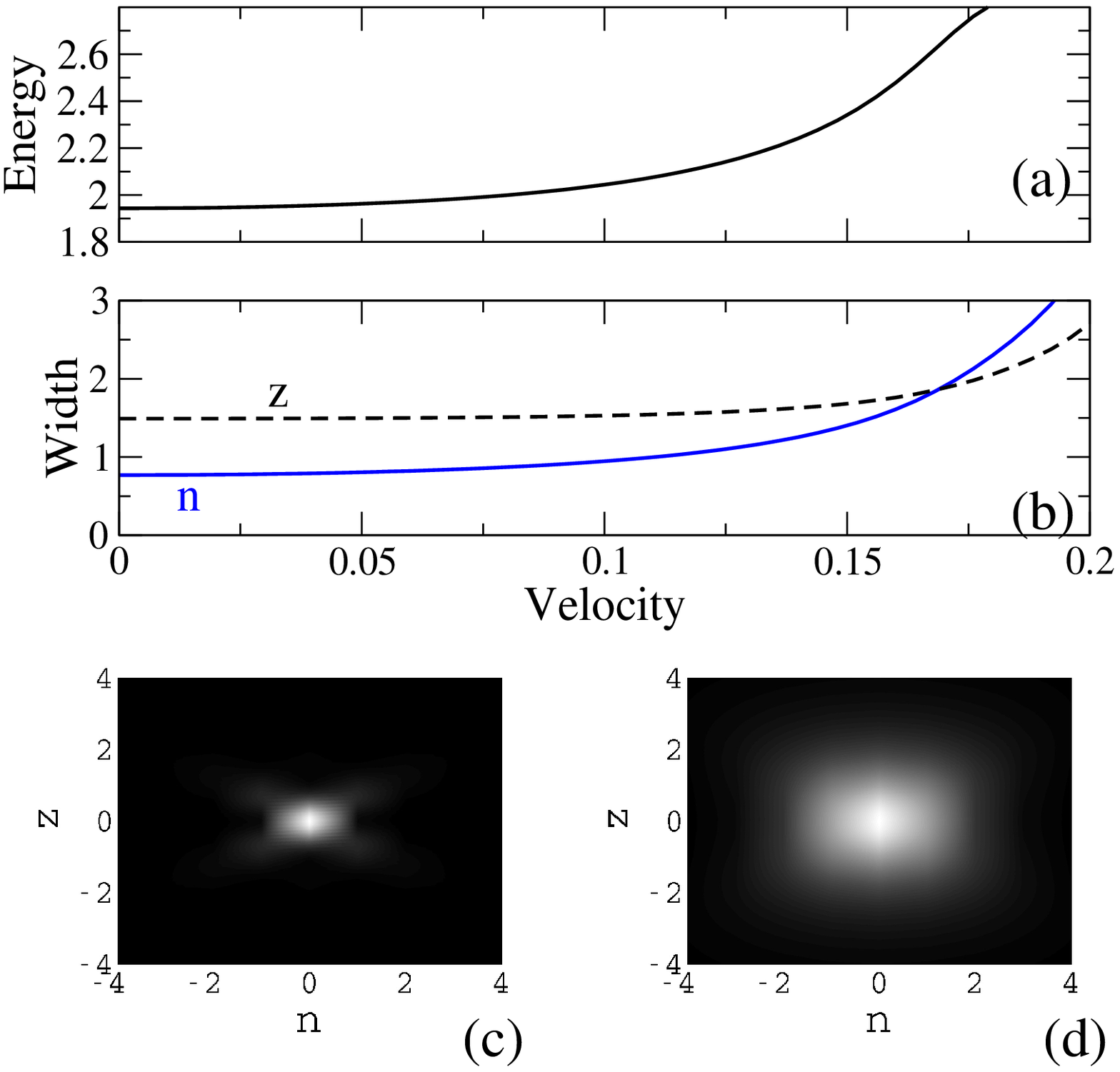}{familyVelocity}{ (colour online) Moving
slow-light optical bullets with the frequency detuning $\omega=0.5$.
Notations are the same as in Fig.~\rpict{familyFreq}, and the
velocity is normalized to the speed of light away from the Bragg
resonance. Profiles are shown for the velocities (c)~$V=0.1$ and
(d)~$V=0.2$. }

We find that the nature of wave dispersion is fundamentally altered
for the waveguide structure with {\em out-of-phase shift of the
neighboring gratings} [i.e., when $\varphi=\pi$; see
Fig.~\rpict{array}(b)]. The corresponding dispersion relation has a
different form, $\omega(K,\beta) = \pm \{\rho^2+[\beta-2 C
\cos(K)]^2\}^{1/2}$. Moreover, for any propagation angle defined by
the transverse Bloch wavevector component $K$, the width and
position of the one-dimensional frequency gap remains the same,
$|\omega| < \rho$. This unusual property leads to remarkable
spectral features. First, the 2D (quasi-)gap is always present in
the spectrum irrespectively to the grating strength ($\rho$) and
coupling between the waveguides ($C$). Second, the shape of
isofrequency contours does not depend on frequency in the
transmission band, see Fig.~\rpict{array}(d). This means that the
beam diffraction remains the same even when the band edge is
approached.

The dependence of diffraction on frequency strongly affects the
shaping of short pulses with broad frequency spectra. We perform
numerical modeling of the pulse dynamics, when the input beam is
focused into a single waveguide. The pulse duration is $T = 10$, and
the central frequency is $\omega_0 = 1$. The snapshots of the light
intensity after the pulse enters the photonic lattice are shown in
Figs.~\rpict{array}(e,f). For the structure with the in-phase
gratings [Figs.~\rpict{array}(e)], the spatial diffraction strongly
depends on frequency, creating a colored pattern similar to the
superprism effect in photonic crystals~\cite{Kosaka:1998-10096:PRB}.
For the phase-shifted gratings, all the components inside the
frequency band experience exactly the same diffraction and propagate
together. As a result, the intensity profile
[Figs.~\rpict{array}(e)] has precisely the same shape as for the
discrete diffraction of monochromatic
beams~\cite{Christodoulides:2003-817:NAT}, with well-pronounced
field minima and peak intensities at the outer wings.

The unique features of linear spectrum in arrays with phase-shifted
gratings suggest that these structures provide optimal conditions
for a nonlinear control of the pulse dynamics. In particular, since
the 2D gap appears for any values of the grating strength and
waveguide coupling, it is possible to choose these parameters
independently in order to balance the rates of dispersion and
diffraction. This allows for {\em simultaneous compensation} of the
pulse broadening in space and time through the nonlinear
self-trapping effect. Indeed, we find numerically localized
solutions of Eqs.~\reqt{CM} for stationary and moving light bullets
of the form of solitary waves $\{u,v\}_n(z,t) =
\{\widetilde{u},\widetilde{v}\}_n(z - V t) \exp( - i \omega t )$,
where $V$ is the propagation velocity. We confirm that localization
is possible in both the cases of positive ($\gamma=+1$) and negative
($\gamma=-1$) nonlinear response, since anomalous or normal
dispersion regimes can be accessed on either edges of the photonic
bandgap. The soliton solutions for different signs of $\gamma$ can
be mapped by changing $\omega \rightarrow -\omega$ and making a
corresponding transformation of pulse profiles. In
Fig.~\rpict{familyFreq}, we show the characteristics of immobile
solitons such as energy and width vs. the frequency detuning inside
the bandgap for self-focusing nonlinearity. These solitons have a
well-pronounced X-shape near the upper edge of the gap, becoming
more localized inside the gap as the pulse energy is increased. We
notice however that these are fully localized states that should not
be confused with the so-called X-waves~\cite{Droulias:2005-1827:OE}
which remain quasi-localized only over a finite propagation
distance. The gap solitons can also propagate along the waveguides,
and we present the characteristic of moving solitons in
Fig.~\rpict{familyVelocity}. These slow-light bullets become more
extended in both the transverse and longitudinal directions as the
propagation velocity is increased.

\pict{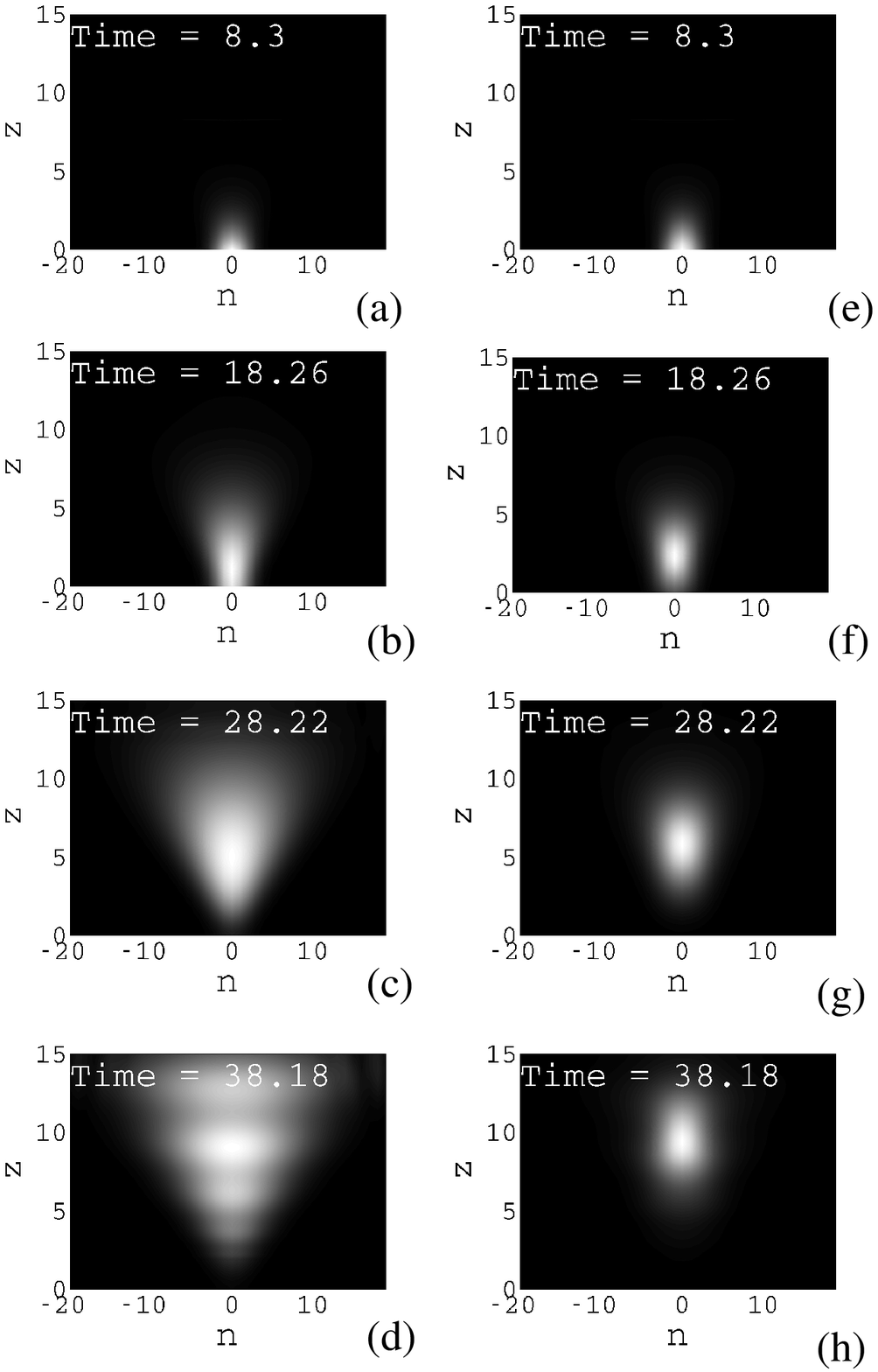}{bpm}{ Snapshots of field intensities for an
optical pulse propagating in a waveguide array structure show in
Fig.~\rpict{array}(d): (a-d)~Linear broadening due to spatial
diffraction and temporal dispersion; (e-h)~Nonlinear self-trapping
in space and time and formation of an optical bullet. }

Finally, we perform numerical simulations of the spatiotemporal
pulse dynamics in these structures. In the linear regime, the pulse
broadens in both transverse and longitudinal directions
[Figs.~\rpict{bpm}(a-d)]. Nonlinear self-action results in the pulse
self-trapping in both space and time [Figs.~\rpict{bpm}(e-h)]. In
this example, the velocity of the generated light bullet is below
30\% of the speed of light in the absence of the Bragg grating, and
smaller velocities can be accessed as well by controlling the
frequency of the input pulse. The profile of the self-trapped state
shown in Fig.~\rpict{bpm}(h) closely resembles the shape of the
exact soliton solution shown in Fig.~\rpict{familyVelocity}(d).

In conclusion, we have revealed that both spatial diffraction and
temporal dispersion can be engineered independently near the edge of
the photonic bandgap in the waveguide arrays with phase-shifted
Bragg gratings. We have shown that such structures possess
quasi-two-dimensional photonic bandgaps and provide optimal
conditions for self-localization of pulses in nonlinear media. We
have demonstrated that these waveguide arrays can be employed for
shaping and control of optical pulses simultaneously in space and
time, and allow the formation of optical bullets propagating with
slow group velocities. Such slow-light optical bullets offer novel
possibilities for all-optical switching, steering, and control of
short pulses.

We thanks M. de Sterke  and B.~Eggleton for useful discussions. This
work has been supported by the Australian Research Council.

\end{sloppy}

\begin{thebibliography}{10}

\bibitem{Soljacic:2002-2052:JOSB}
M. Soljacic and J.D. Joannopoulos, Nature Materials {\bf 3} 211
(2004).

\bibitem{Eggleton:1999-587:JOSB}
B.~J. Eggleton, C.~M. {de Sterke}, and R.~E. Slusher, J. Opt. Soc.
Am. B {\bf 16}, 587 (1999).

\bibitem{Millar:1999-685:OL}
P. Millar, R.~M. {De la Rue}, T.~F. Krauss, J.~S. Aitchison,
N.~G.~R. Broderick, and D.~J. Richardson, Opt. Lett. {\bf 24}, 685
(1999).

\bibitem{Vlasov:2005-65:NAT}
Y.~A. Vlasov, M. O'Boyle, H.~F. Hamann, and S.~J. McNab, Nature {\bf
438}, 65 (2005).

\bibitem{Gersen:2005-73903:PRL}
H. Gersen, T.~J. Karle, R.~J.~P. Engelen, W. Bogaerts, J.~P.
Korterik, N.~F. Hulst, van, T.~F. Krauss, and L. Kuipers, Phys. Rev.
Lett. {\bf 94}, 073903 (2005).

\bibitem{Jacobsen:2005-7861:OE}
R.~S. Jacobsen, A.~V. Lavrinenko, L.~H. Frandsen, C. Peucheret, B.
Zsigri, G. Moulin, J. Fage~Pedersen, and P.~I. Borel, Opt. Express
{\bf 13}, 7861 (2005).

\bibitem{Kivshar:2003:OpticalSolitons}
Yu.~S. Kivshar and G.~P. Agrawal, {\em {Optical Solitons: From
Fibers to Photonic Crystals}} (Academic Press, San Diego, 2003).

\bibitem{Christodoulides:2003-817:NAT}
D.~N. Christodoulides, F. Lederer, and Y. Silberberg, Nature {\bf
424}, 817 (2003).

\bibitem{Fleischer:2005-1780:OE}
J.~W. Fleischer, G. Bartal, O. Cohen, T. Schwartz, O. Manela, B.
Freedman, M. Segev, H. Buljan, and N.~K. Efremidis, Opt. Express
{\bf 13}, 1780 (2005).

\bibitem{Rosberg:2005-2293:OL}
C.~R. Rosberg, D.~N. Neshev, A.~A. Sukhorukov, Yu.~S. Kivshar, and
W. Krolikowski, Opt. Lett. {\bf 30}, 2293 (2005).

\bibitem{Eisenberg:2001-43902:PRL}
H.~S. Eisenberg, R. Morandotti, Y. Silberberg, S. Bar~Ad, D. Ross,
and J.~S. Aitchison, Phys. Rev. Lett. {\bf 87}, 043902 (2001).

\bibitem{Silberberg:1990-1282:OL}
Y. Silberberg, Opt. Lett. {\bf 15}, 1282 (1990).

\bibitem{Altug:2005-111102:APL}
H. Altug and J. Vuckovic, Appl. Phys. Lett. {\bf 86}, 111102 (2005).

\bibitem{Joannopoulos:1995:PhotonicCrystals}
J.~D. Joannopoulos, R.~D. Meade, and J.~N. Winn, {\em Photonic
Crystals: Molding the Flow of Light} (Princeton University Press,
Princeton, 1995).

\bibitem{Dohnal:2005-209:STAM}
T. Dohnal and A.~B. Aceves, Stud. Appl. Math. {\bf 115}, 209 (2005).

\bibitem{Agrawal:1989:NonlinearFiber}
G.~P. Agrawal, {\em {Nonlinear Fiber Optics}} (Academic Press, New
York, 1988).

\bibitem{Mak:2004-66610:PRE}
W.~C.~K. Mak, B.~A. Malomed, and P.~L. Chu, Phys. Rev. E {\bf 69},
066610 (2004).

\bibitem{Gubeskys:2004-283:EPD}
A. Gubeskys and B.~A. Malomed, Eur. Phys. J. D {\bf 28}, 283 (2004).

\bibitem{Kosaka:1998-10096:PRB}
H. Kosaka, T. Kawashima, A. Tomita, M. Notomi, T. Tamamura, T. Sato,
and S. Kawakami, Phys. Rev. B {\bf 58}, R10096 (1998).

\bibitem{Droulias:2005-1827:OE}
S. Droulias, K. Hizanidis, J. Meier, and D.~N. Christodoulides, Opt.
Express {\bf 13}, 1827 (2005).

\end{thebibliography}
\end{document}